\documentclass[12pt, fleqn]{article}
\usepackage[cp1251]{inputenc}
\usepackage{latexsym,amsfonts,amssymb}
\usepackage{graphicx}
\usepackage{amsbsy}
\usepackage{amsmath}
\usepackage{epsf}
\usepackage{cite}

\newcommand{\be} {\begin{equation}}
\newcommand{\ee} {\end{equation}}
\newcommand{\ba}{\begin{array}{l}}
\newcommand{\ea}{\end{array}}
\newcommand{\ts}{\textstyle}  %@@

\newtheorem{theo} {Theorem}% [] %chapter]  %@@
\newtheorem{remark}{Remark}
\newcommand{\bteo}{\begin{theo}}  %@@
\newcommand{\et}{\end{theo}}

\newcommand{\pt} {\partial}
\newcommand{\al} {\alpha}

\newcommand{\bn}{\begin{note}}
\newcommand{\en}{\end{note}}

\begin{document}

\begin{center}
{\large \bf Lie symmetries of the Shigesada--Kawasaki--Teramoto
system}
\medskip\\
{\bf Roman CHERNIHA~$^{\dag}$\footnote{\small  Corresponding author. E-mails: r.m.cherniha@gmail.com; cherniha@gmail.com}, Vasyl' DAVYDOVYCH~$^{\dag}$ and  Liliia MUZYKA~$^\ddag$  } \\
%% \footnote{\small e-mail: cherniha@imath.kiev.ua }\\
\medskip
{\it $^\dag$~Institute of Mathematics,  NAS
of Ukraine,\\ 3 Tereshchenkivs'ka Street, 01601 Kyiv, Ukraine}\\
{\it $^\ddag$~Faculty of Information Systems, Physics and Mathematics,\\
Lesya Ukrainka Eastern European National University, \\
13 Voly Avenue,  43025 Lutsk, Ukraine} \\
 % E-mail: Liliia\underline{  }Myroniuk@univer.lutsk.ua\\
\end{center}

\begin{abstract}

The  Shigesada--Kawasaki--Teramoto system, which consists  of two
reaction-diffusion   equations with variable cross-diffusion and
quadratic nonlinearities,  is considered. The system is the most
important  case of  the  biologically motivated   model proposed by
Shigesada \emph{et al.} (\emph{J. Theor. Biol.} \textbf{79}(1979)
83--99). A complete description of Lie symmetries for this system is
derived. It is proved that the Shigesada--Kawasaki--Teramoto system
admits a wide range of different  Lie symmetries depending on
coefficient values. In particular, the Lie symmetry operators with
highly unusual structure are unveiled
%uncovered
 and applied for finding exact
solutions of the relevant nonlinear system with  cross-diffusion.

\end{abstract}

\textbf{Keywords:} reaction-diffusion system; cross-diffusion; Lie symmetry; exact solution.

\newpage

\section{\bf Introduction} \label{sec-1}

In 1979  Shigesada \emph{et al.} \cite{sh-ka-te} proposed a
mathematical model to describe the densities of two biological
species, which takes into
 account
%the pressures created by two mutually competing species.
the heterogeneity of the environment and nonlinear dispersive
movements of the individuals of these populations. The model was
developed on the basis of Morisita's phenomelogical theory of
environmental density and  has the form
 \be \label {1}
\ba
 \medskip
u_t = [(d_1+d_{11}u+d_{12}v) u]_{xx} +(W_xu)_x+u(a_1  - b_1 u- c_1v), \\
v_t = [(d_2+d_{21}u+d_{22}v) v]_{xx} +(W_xv)_x+v(a_2  - b_2 u- c_2v),
\ea \ee
where the functions $u$ and  $v$
arising in system (\ref{1})
give the densities
%%(concentrations)
 of two competing species
in space
%%(the variable $x$)
 and time,
%% (the variable $t$),
$d_1$ and  $d_2$ denote the diffusion coefficients, $d_{12}v$ and
$d_{21}u$ are so-called cross-diffusion pressures, $d_{11}u$ and
$d_{22}v$ are  intra-diffusion pressures, $a_1$ and $a_2$ are  the
intrinsic growth coefficients, $b_1$ and $c_2$ denote the
coefficients of intra-specific competitions, $b_2$ and $c_1$ denote
the coefficients of inter-specific competitions. The function $W(x)$
is so-called environmental potential, which is assumed to be known.
 Nevertheless, the authors of \cite{sh-ka-te} assumed that
the  environmental potential may be a non-constant function, system
(\ref{1}) with $W(x)=const$ is usually referred as
Shigesada--Kawasaki--Teramoto (SKT) system (model).
It was shown  by numerical simulations  that system (\ref{1}) possesses
solutions describing coexistence of the species by the spatial segregation of habitat \cite{sh-ka-te}.
This kind of coexistence results from the mutual interferences of the species
and the heterogeneity of the environment and means  a steady-state segregation of  densities of two competing species.
The existence of  the steady-state segregation clearly depends on the initial distributions of $u$ and  $v$ and the parameter values of
(\ref{1}).

It is worth to note that  the SKT system  with $d_{ij}=0 \ (i=1,2,
\, j=1,2)$ and $W(x)=const$ reduces to the well-known diffusive
Lotka--Volterra (DLV) system

\be \label {2a}
\ba
 \medskip
u_t = d_1u_{xx}+u(a_1  - b_1 u- c_1v), \\
v_t = d_2v_{xx} +v(a_2  - b_2 u- c_2v). \ea \ee

 Starting from the pioneer works   \cite{co-smo, brown},
the  conditions of  existence, uniqueness and global stability/instability
of solutions for the DLV system  and the SKT system
 were  investigated  extensively  by many
scholars
%%(see  the extensive list of  references in
(see \cite{mimura-et-all-84, lou-ni,shim-03,wu-05,kuto,des-tre-2015,lou-win-15}  and the papers cited therein).
%Nevertheless
Although there are not many papers devoted to finding   exact
solutions  (especially in explicit form) of these nonlinear systems.
To the best of our knowledge, there are only a few papers, in which exact solutions of the SKT
 system  were found \cite{ch-2001,horstmann-07,ch-my2008}. The list of such references for
  the DLV systems  is wider
 \cite{rod-mimura-2000,ch-du-2004,3-ch-dav-2011, 3-hung2011, ch-dav2013, 3-hung2012}
 and the results are summarized in the very recent book \cite{ch-dav-2016}.
  Note
 that a wide range of exact solutions of the Lotka--Volterra type systems with
 power diffusivities (without cross-diffusion) were constructed in
 in \cite{ch-king4,ch-myr-2006} (see also \cite{ch-dav-2016} and references cited therein).

In this paper system (\ref{1}) with $W(x)=const$, i.e., the SKT
system \be\label {2}
 \ba
 \medskip
 u_t = [(d_1+d_{11}u+d_{12}v)u]_{xx}+u(a_1-b_1u-c_1v),\\
 v_t = [(d_2+d_{21}u+d_{22}v)v]_{xx}+v(a_2-b_2u-c_2v)
 \ea
  \ee
  is under study.
 %% and can be referred as  the  generalized Shigesada-Kawasaki-Teramoto system (SKT system).
 Depending on signs of the parameters $a_k$, $b_k$ and
  $c_k$ ($k=1,2$) the SKT system (\ref{2}) can describe different
  types of species interactions (competition, mutualism and
  prey--predator interaction). Hereafter  $d_{ij} \ (i=1,2, \ j=1,2)$ are assumed to be real constants and
  $ d^2_{12}+d^2_{21}\not=0$, i.e., we consider only the systems with cross-diffusion. Moreover, we
   always assume that both equations involve diffusion, i.e., both equations are the second-order PDEs.

A complete description of Lie symmetries for this system with
$d_{ij}=0 \  (i=1,2, \ j=1,2)$, i. e., DLV system, is derived in
\cite{ch-du-2004} and extended on the case of three-component DLV
systems in \cite{ch-dav2013}.

It should be noted that  the system (\ref{2}) with $d_{12}=d_{21}=0$
is a particular case of the general reaction-diffusion (RD) system
with variable diffusivities.
 Lie symmetry of such system was completely described  in \cite{ch-king4,2-knyazeva-94}.
%So we do not investigate these cases below.

The paper is organized as follows. In Section~\ref{sec-2},
 a problem of the Lie symmetry
  classification of system (\ref{2}) is completely solved.
   It is proved that the SKT system  admits a wide range
   of different  Lie symmetries depending on coefficient
    values. In particular, the Lie symmetry operators with
     highly unusual structure are unveiled. These operators are
     nonlinear with respect to the dependent variables $u$ and $v$,
     hence the do not occur in RD systems without
      cross-diffusion.  In Section~\ref{sec-3}, we present some examples of
      exact solutions and their possible interpretation.  Finally,
       we  summarize  the results obtained and  present some conclusions  in
the last section.

\section{\bf Lie symmetry classification of the SKT system (\ref{2})
} \label{sec-2}

It is easily checked that  the system (\ref{2}) with arbitrary coefficients is invariant under the operators
 \be\label{3}
P_x =\frac{\ts\pt}{\ts \pt x}\equiv \pt_x,
 \, P_t = \frac{\ts\pt}{\ts \pt t}\equiv \pt_t.
 \ee
Hereafter we  call this algebra the trivial Lie algebra (other terminology used for this algebra is the
 `principal algebra'  and the  `kernel of maximal invariance algebras')
%`kernel of the Lie algebras'  in this context)
of the SKT system (\ref{2}).
Thus, we aim to find all coefficients arising in  the nonlinear system (\ref{2}) that lead to extensions of its trivial Lie algebra (\ref{3}). Because the SKT system (\ref{2}) contains 12 parameters it is a non-trivial task
and the result obtained is highly non-trivial.

Taking into account the known results for the DLV system
\cite{ch-du-2004},  we consider only the systems with
cross-diffusion ($ d^2_{12}+d^2_{21}\not=0$), which  cannot be
reduced to the systems without cross diffusion  ($d_{12}=d_{21}=0$)
by any point (local) transformations.

%We also note that  there is a  special case when the SKT system (\ref{2}) involves a  %first-order equation (i.e., ODE with the variable $x$ as a parameter) or can be %reduced to a such system.
% Because system (\ref{2}) with   $d_{2}=d_{21}=d_{22}=0$  can
%describe interactions  involving species (cells) $V$, which are not
%able to diffuse in space (a typical example is plankton), we examine this %case separately.

It is convenient to separate   the results obtained into three parts depending on the coefficients arising in the SKT system (\ref{2}). Thus,
three theorems will be presented for the following cases:

\begin{enumerate}
\item  System (\ref{2}) with both standard diffusion  and cross-diffusion, and non-vanish reaction term(s).

    \item  System (\ref{2}) with  cross-diffusion diffusion only, i.e., $d_1=d_2=d_{11}=d_{22}=0$, and non-vanish reaction term(s).

        \item  System (\ref{2}) with both standard diffusion and  cross-diffusion, however   reaction terms  vanish, i.e., $a_k=b_k=c_k=0 \ (k=1,2).$

\end{enumerate}

Notably that  the nonlinear system (\ref{2}) without reaction terms
may
describe the crystal growth processes (see  \cite{fe-ros-89,ch-fe-98} for details). %We also  examine this case separately.

Our main result can be formulated in form of
Theorems~\ref{th-1}--\ref{th-3} presenting  the  complete Lie
symmetry classification of system (\ref{2}) in each of the cases
listed above.

\begin{theo}\label{th-1}
All possible  maximal algebras of invariance (up to equivalent
representations generated by transformations of the form (\ref{4}))
of the SKT system (\ref{2}) are presented in  Table~\ref{tab-1}. Any
other system of the form (\ref{2}) which is invariant with respect
to (w.r.t.) the three- and higher-dimensional maximal algebra of
invariance (MAI)   is reduced by a substitution of the form
\be\label{4} \ba
\medskip
t^*=\al_{00}e^{\al_0 t}, \quad  x^*= \al_{01}x, \\
u^*=\al_{10}+\al_{11}u+\al_{12}v+\al_{13}e^{\al_1t}u, \\
v^*=\al_{20}+\al_{21}u+\al_{22}v+\al_{23}e^{\al_2t}v \ea \ee to one
of those given in Table~\ref{tab-1} (constants $\al_{ij}$ are
determined by the form of the system in question).
\end{theo}

\textbf{The sketch of the proof.}  We can rewrite the system (\ref{2}) in equivalent form:
\be\label{5} \ba
\medskip
u_t=d_1u_{xx}+2d_{11}uu_{xx}+d_{12}vu_{xx}+d_{12}uv_{xx}+\\
\qquad2d_{11}{u_{x}}^2+2d_{12}u_{x}v_{x} +a_1u-b_1u^2-c_1uv,\\
v_t=d_2v_{xx}+2d_{22}vv_{xx}+d_{21}uv_{xx}+d_{21}vu_{xx}+\\
\qquad2d_{22}{v_{x}}^2+2d_{21}u_{x}v_{x}+a_2v-c_2v^2-b_2uv.\\
 \ea
  \ee
\quad The proof of the theorem is based
 on the classical Lie scheme (see, e.g.,  \cite{bl-anco, fss,
ol}); however, it is
 highly  non-trivial and
cumbersome because  the system (\ref{2}) contains twelve arbitrary
coefficients. Here we give  an outline of how the proof proceeds.
According to the Lie approach, the system (\ref{2})
is considered as a manifold $\cal
M$$=\{S_1=0, S_2=0\}$:
\be\label{6} \ba
\medskip
S_1\equiv
-u_t+d_1u_{xx}+2d_{11}uu_{xx}+d_{12}vu_{xx}+d_{12}uv_{xx}+\\
\qquad2d_{11}{u_{x}}^2+2d_{12}u_{x}v_{x} +a_1u-b_1u^2-c_1uv=0,\\
S_2\equiv
-v_t+d_2v_{xx}+2d_{22}vv_{xx}+d_{21}uv_{xx}+d_{21}vu_{xx}+\\
\qquad2d_{22}{v_{x}}^2+2d_{21}u_{x}v_{x}+a_2v-c_2v^2-b_2uv=0\\
 \ea
  \ee
in the space of the following variables:  %@@
$$t, x, u, v, u_t, v_t,u_{x},v_{x}, u_{xx},v_{xx},   $$
where subscripts  %@@
 $t$ and  $x $  to the functions $u$ and $v$  %@@
denote differentiation with respect to these variables.  %@@

System (\ref{2}) is invariant under the transformations generated by the  %@@
infinitesimal operator
\be\label{7}
X=\xi^0(t,x,u,v)\pt_t+\xi^1(t,x,u,v)\pt_x+\eta^1(t,x,u,v)\pt_u+\eta^2(t,x,u,v)\pt_v,
\ee when the following invariance conditions are satisfied:
\be\label{8}
\mbox{\raisebox{-1.4ex}{$\stackrel{\displaystyle  %@@
X}{\scriptstyle 2}$}}S_1\vert_{\cal M}=0,
\mbox{\raisebox{-1.4ex}{$\stackrel{\displaystyle  %@@
X}{\scriptstyle 2}$}}S_2\vert_{\cal M}=0. \ee
The operator $ \mbox{\raisebox{-1.4ex}{$\stackrel{\displaystyle  %@@
X}{\scriptstyle 2}$}} $   is the second  %@@
 prolongation of the operator $X$, i.e.,  %@@
\be\label{9}  %@@
\ba
 \medskip
\mbox{\raisebox{-1.4ex}{$\stackrel{\displaystyle  %@@
X}{\scriptstyle 2}$}} =X+\rho^1_t\frac{\pt}{\pt
u_t}+\rho^2_t\frac{\pt}{\pt v_t}+\rho^1_x\frac{\pt}{\pt
u_x}+\rho^2_x\frac{\pt}{\pt v_x}+  \sigma^1_{tx}\frac{\pt}{\pt
u_{tx}}+\sigma^2_{tx}\frac{\pt}{\pt v_{tx}} +\\ \qquad
\sigma^1_{tt}\frac{\pt}{\pt u_{tt}}+\sigma^2_{tt}\frac{\pt}{\pt
v_{tt}} +\sigma^1_{xx}\frac{\pt}{\pt
u_{xx}}+\sigma^2_{xx}\frac{\pt}{\pt v_{xx}},
\ea  %@@
\ee  %@@
where the coefficients $\rho$ and $\sigma$ with relevant subscripts  %@@
are calculated by well-known formulae   %@@
(see, e.g., \cite{fss}).  %@@

Substituting (\ref{9})  into (\ref{8}) and eliminating the derivatives
$ u_t$ and $ v_t$ using (\ref{6}),  we can split this relation into  %@@
separate parts for the derivatives  $u_x, v_x, u_{xx}, v_{xx}, u_x v_x$.
Finally, after the
relevant calculations, we obtain the following system
of determining equations (DEs):
\be\label{10}
\xi^0_x=\xi^0_u=\xi^0_v=\xi^1_u=\xi^1_v=0,\\
\ee \be\label{11}
d_{21}v\eta_{uu}^1+(d_2+d_{21}u+2d_{22}v)\eta_{uu}^2+2(d_{21}-d_{11})\eta_u^2=0,\\
\ee \be\label{12}
(d_1+2d_{11}u+d_{12}v)\eta_{uu}^1+d_{12}u\eta_{uu}^2+
2d_{11}(\eta_u^1+\xi_t^0-2\xi_x^1)+2d_{12}\eta_u^2=0,\\
\ee \be\label{13}
(d_1+2d_{11}u+d_{12}v)\eta_{vv}^1+d_{12}u\eta_{vv}^2+2(d_{12}-d_{22})\eta_v^1=0,
\ee  \be\label{14}
d_{21}v\eta_{vv}^1+(d_2+d_{21}u+2d_{22}v)\eta_{vv}^2+2d_{21}\eta_v^1+
2d_{22}(\eta_{v}^2+\xi_t^0-2\xi_x^1)=0,\\
\ee
 \be\label{15}
(d_1+2d_{11}u+d_{12}v)\eta_{uv}^1+d_{12}u\eta_{uv}^2+(2d_{11}-d_{21})\eta_v^1+
d_{12}(\eta_{v}^2+\xi_t^0-2\xi_x^1)=0,\\
\ee \be\label{16}
d_{21}v\eta_{uv}^1+(d_2+d_{21}u+2d_{22}v)\eta_{uv}^2+d_{21}(\eta_u^1+\xi_t^0-2\xi_x^1)+
(2d_{22}-d_{12})\eta_{u}^2=0,\\
\ee \be\label{17} \ba   \medskip
d_{12}u\eta_u^1-(d_1-d_2+(2d_{11}-d_{21})u+(d_{12}-2d_{22})v)\eta_v^1-d_{12}u\eta_v^2-\\
d_{12}\eta^1-d_{12}u(\xi_t^0-2\xi_x^1)=0,\\
\ea \ee \be\label{18}\ba   \medskip
d_{21}v\eta_u^1-(d_1-d_2+(2d_{11}-d_{21})u+(d_{12}-2d_{22})v)\eta_u^2-d_{21}v\eta_v^2+\\
d_{21}\eta^2+d_{21}v(\xi_t^0-2\xi_x^1)=0,\\
\ea \ee \be\label{19}
d_{21}v\eta_v^1-d_{12}u\eta_u^2-2d_{11}\eta^1-d_{12}\eta^2-(d_1+2d_{11}u+d_{12}v)(\xi_t^0-2\xi_x^1)=0,\\
\ee \be\label{20} d_{21}v\eta_v^1-d_{12}u\eta_u^2+d_{21}\eta^1+
2d_{22}\eta^2+(d_2+d_{21}u+2d_{22}v)(\xi_t^0-2\xi_x^1)=0,\\
\ee \be\label{21}
2d_{21}v\eta_{xu}^1+2(d_2+d_{21}u+2d_{22}v)\eta_{xu}^2+2d_{21}\eta_x^2-
d_{21}v\xi_{xx}^1=0,\\
\ee \be\label{22}\ba   \medskip
2(d_1+2d_{11}u+d_{12}v)\eta_{xu}^1+2d_{12}u\eta_{xu}^2+4d_{11}\eta_x^1+
2d_{12}\eta_x^2+\\
\xi_t^1-(d_1+2d_{11}u+d_{12}v)\xi_{xx}^1=0,\\
\ea \ee \be\label{23}
2(d_1+2d_{11}u+d_{12}v)\eta_{xv}^1+2d_{12}u\eta_{xv}^2+2d_{12}\eta_x^1-
d_{12}u\xi_{xx}^1=0,\\
\ee \be\label{24}\ba   \medskip
2d_{21}v\eta_{xv}^1+2(d_2+d_{21}u+2d_{22}v)\eta_{xv}^2+2d_{21}\eta_x^1+4d_{22}\eta_x^2+\\
\xi_t^1-(d_2+d_{21}u+2d_{22}v)\xi_{xx}^1=0,\\
\ea \ee \be\label{25} \ba
\medskip
\eta_t^1+(a_1u-b_1u^2-c_1uv)\eta_u^1+(a_2v-b_2uv-c_2v^2)\eta_v^1-\\(a_1-2b_1u-c_1v)\eta^1+c_1u\eta^2-
(d_1+2d_{11}u+d_{12}v)\eta_{xx}^1-d_{12}u\eta_{xx}^2-\\
(a_1u-b_1u^2-c_1uv)\xi_t^0=0,\\
\ea \ee \be\label{26} \ba
\medskip
\eta_t^2+(a_2v-c_2v^2-b_2uv)\eta_v^2+(a_1u-b_1u^2-c_1uv)\eta_u^2+\\b_2v\eta^1-(a_2-b_2u-2c_2v)\eta^2
-d_{21}v\eta_{xx}^1-(d_2+d_{21}u+2d_{22}v)\eta_{xx}^2-\\
(a_2v-b_2uv-c_2v^2)\xi_t^0=0. \ea \ee

The system of DEs (\ref{10})---(\ref{26}) is very cumbersome and one
needs to establish how the functions $\eta^1$  and $\eta^2$  depend
on the variables  $u$ and $v$. It is well-known that this dependence
is linear in the case of RD equations \cite{dor1} and systems
\cite{ch-king, ch-king03, ch-king4}. It turns out that the
cross-diffusion terms in RD systems may lead to
 a completely different result. In order to prove this, we need to examine differential
 consequences of equations
(\ref{17})---(\ref{20}) w.r.t. $u$ and $v$: \be\label{27} \ba
\medskip
d_{12}u\eta^1_{uu}-(d_1-d_2+(2d_{11}-d_{21})u+(d_{12}-2d_{22})v)\eta^1_{uv}-d_{12}u\eta^2_{uv}-\\
(2d_{11}-d_{21})\eta^1_v-d_{12}\eta^2_v-d_{12}(\xi^0_t-2\xi^1_x)=0,
\ea \ee \be\label{28} \ba
\medskip d_{12}u\eta^1_{uv}-(d_1-d_2+(2d_{11}-d_{21})u+(d_{12}-2d_{22})v)\eta^1_{vv}-d_{12}u\eta^2_{vv}+\\
2(d_{22}-d_{12})\eta^1_v=0, \ea \ee \be\label{29} \ba
\medskip
d_{21}v\eta^1_{uu}-(d_1-d_2+(2d_{11}-d_{21})u+(d_{12}-2d_{22})v)\eta^2_{uu}-d_{21}v\eta^2_{uv}+\\
2(d_{21}-d_{11})\eta^2_u=0, \ea \ee \be\label{30} \ba
\medskip
d_{21}v\eta^1_{uv}-(d_1-d_2+(2d_{11}-d_{21})u+(d_{12}-2d_{22})v)\eta^2_{uv}-d_{21}v\eta^2_{vv}+\\
d_{21}\eta^1_u-(d_{12}-2d_{22})\eta^2_u+d_{21}(\xi^0_t-2\xi^1_x)=0,\ea
\ee \be\label{31} \ba
\medskip
d_{21}v\eta^1_{uv}-d_{12}u\eta^2_{uu}-2d_{11}\eta^1_u-2d_{12}\eta^2_u-2d_{11}(\xi^0_t-2\xi^1_x)=0,
\ea \ee \be\label{32} \ba
\medskip
d_{21}v\eta^1_{vv}-d_{12}u\eta^2_{uv}-(2d_{11}-d_{21})\eta^1_v-d_{12}\eta^2_v-d_{12}(\xi^0_t-2\xi^1_x)=0,
\ea \ee \be\label{33} \ba
\medskip
d_{21}v\eta^1_{uv}-d_{12}u\eta^2_{uu}+d_{21}\eta^1_u-(d_{12}-2d_{22})\eta^2_u+d_{21}(\xi^0_t-2\xi^1_x)=0,
\ea \ee \be\label{34} \ba
\medskip
d_{21}v\eta^1_{vv}-d_{12}u\eta^2_{uv}+2d_{21}\eta^1_v+2d_{22}\eta^2_v+2d_{22}(\xi^0_t-2\xi^1_x)=0.
\ea \ee

The next crucial step is to remove all the first-order derivatives
from (\ref{27})---(\ref{34}) using the system of DEs
(\ref{10})---(\ref{26}). In fact, if one subtracts equation
(\ref{29}) from (\ref{11}), equation  (\ref{34}) from (\ref{14}),
equation  (\ref{30}) from (\ref{16}), equation  (\ref{33}) from
(\ref{16})  and adds (\ref{12}) to (\ref{31}), (\ref{13}) to
(\ref{28}), (\ref{15}) to (\ref{27}), (\ref{15}) to (\ref{32}) then
the linear algebraic system
\[\ba (d_1+2d_{11}u+d_{12}v)\eta^1_{uu}+d_{21}v\eta^1_{uv}=0,\\
d_{12}u\eta^1_{uu}+(d_2+d_{21}u+2d_{22}v)\eta^1_{uv}=0,\\
d_{12}u\eta^1_{uv}+(d_2+d_{21}u+2d_{22}v)\eta^1_{vv}=0,\\
(d_1+2d_{11}u+d_{12}v)\eta^1_{uv}+d_{21}v\eta^1_{vv}=0,\\
(d_1+2d_{11}u+d_{12}v)\eta^2_{uu}+d_{21}v\eta^2_{uv}=0,\\
d_{12}u\eta^2_{uu}+(d_2+d_{21}u+2d_{22}v)\eta^2_{uv}=0,\\
d_{12}u\eta^2_{uv}+(d_2+d_{21}u+2d_{22}v)\eta^2_{vv}=0,\\
(d_1+2d_{11}u+d_{12}v)\eta^2_{uv}+d_{21}v\eta^2_{vv}=0\ea\]
to find the function $\eta^1_{uu}$, $\eta^1_{uv}$, $\eta^1_{vv}$,
$\eta^2_{uu}$, $\eta^2_{uv}$ and  $\eta^2_{vv}$ is obtained.
Because this system is overdetermined (8 equations for 6 functions), we have found that its solution only  is
 \be\label{35}
\eta_{uu}^1=\eta_{uv}^1=\eta_{vv}^1=\eta_{uu}^2=\eta_{uv}^2=\eta_{vv}^2=0,\\
\ee
provided $d_1^2+d_2^2+d_{11}^2+d_{22}^2
\neq 0$.

 The case   $d_1=d_2=d_{11}=d_{22}=0$ is special and will be examined separately.

 Thus, solving equations (\ref{10}) and  (\ref{35}), one immediately obtains
\be\label{36} \ba
\medskip
\xi^0=\xi^0(t)$, \quad $\xi^1=\xi^1(t,x),\\
\eta^1=r^1(t,x)u+q^1(t,x)v+p^1(t,x), \\
\eta^2=q^2(t,x)u+r^2(t,x)v+p^2(t,x),\\
\ea \ee
where  $\xi^0$, $\xi^1$, $r^1(t,x)$, $q^1(t,x)$, $p^1(t,x)$,
$q^2(t,x)$, $r^2(t,x)$ and  $p^2(t,x)$  are arbitrary smooth functions at the moment.
Taking into account (\ref{36}), equations
 (\ref{11}) and (\ref{13}) from the system of DEs, are simplified:
 \be\label{37}
[d_{21}-d_{11}]q^2(t,x)=0,\\
\ee \be\label{38}
[d_{12}-d_{22}]q^1(t,x)=0.\\
\ee
Finally, two  different cases should be examined\\
\textbf{1.} \quad $d_{22}\neq d_{12}$. \qquad \qquad \textbf{2.}
\quad  $d_{22}=d_{12}, d_{21}=d_{11}$. \\The third possible case
$d_{21}\neq d_{11}$ is reduced to the first case by the substitution
$u$ $\rightarrow$ $v$, $v$ $\rightarrow$ $u$ and the relevant
renaming.

We do not present further calculations because it is rather a
standard routine to find the coefficients of  the infinitesimal
operator (\ref{7}) provided they have the form (\ref{36}). The
detailed analysis of case  {\bf 1} shows that the systems and MAIs
listed in cases \textbf{3}---\textbf{7}, \textbf{10}---\textbf{12}
and  \textbf{15} of Table~\ref{tab-1} are obtained, while the same
routine for case  {\bf 2} produces the systems and MAIs  listed in
cases
 \textbf{1}, \, \textbf{2}, \,
\textbf{8--9}, \, \textbf{13--14}  and  \textbf{16}.

Finally, we note that each SKT systems listed in Table~\ref{tab-1}
is
 a representative of some other systems, which are reduced to one
 by the point (local)  substitutions indicated in the last column of Table~\ref{tab-1}.
 The explicit forms of these substitutions are

\be\label{37a}\ba
1. \quad  u^*=v, \quad v^*=u.\\
2. \quad u^*=e_{1}u, \quad v^*= e_{2}v.\\
3. \quad u^*=u+\frac{d_1}{2d_{11}}, \quad v^*=v.\\
4. \quad u^*=u, \quad v^*=v+\frac{d_1}{d_{12}}.\\
5. \quad u^*=u+\frac{d_1-d_2}{d_{11}}, \quad v^*=v.\\
6. \quad u^*=u, \quad v^*=v+\frac{d_2-d_1}{d_{12}}.\\
7.  \quad u^*=u, \quad  v^*=bu+cv.\\
8.  \quad u^*=u, \quad  v^*=d_{11}u+d_{12}v.\\
9. \quad u^*=bu+cv, \quad  v^*=d_{11}u+d_{12}v.\\
10. \quad t^*=\frac{1}{a}e^{a t}, \quad u^*=e^{-a t}u, \quad
v^*=e^{-a t}v.\\
11. \quad u^*=ue^{-a_1t}, v^*=v. \\
12. \quad u^*=u, v^*=ve^{-a_2t}. \\
 \ea \ee

 It can be easily seen  that all the substitutions listed above can be united in the form  (\ref{4})
 (of course, constants $\al_{ij}$
must be correctly-specified  by  the system in question).

 The sketch of the proof is now completed. \ $\Box$

\begin{table}
\caption{MAIs of the SKT system (\ref{2}) } \label{tab-1}
\begin{center}
\begin{tabular}{|c|c|c|c|c|}
\hline
\small  &\small {Systems}  & \small {Restrictions} & \small {Basic operators  } & \small {Substitution}   \\
 & &  & \small {of MAI} &  \small {from (\ref{37a})}   \\
\hline
% &&& \\
\small {1.} & \small {$u_t = [(u+\frac{c}{b}v)u]_{xx} +u(a-bu-cv)$}
  & \small {$abc\neq0$} & \small {$P_t, \, P_x, \, Q_1$} & \small {1, 2}    \\
 & \small {$v_t = [(u+\frac{c}{b}v)v]_{xx}+v(2a-bu-cv)$} &  &  & \\
%&&& \\
\hline
% &&& \\
\small {2.} & \small { $ u_t =
[(1+\frac{b}{a}u+\frac{c}{a}v)u]_{xx}
+u(a-bu-cv)$} & \small {$abc\neq0$}  & \small{ $P_t, \, P_x, \, Q_2$} & \small {1, 2}    \\
 & \small {$v_t = [(2+\frac{b}{a}u+\frac{c}{a}v)v]_{xx}
-v(a+bu+cv)$} &  &  & \\
%&&& \\
\hline
% &&& \\
\small {3.} &  \small {$ u_t = [(d_{11}u+d_{12}v)u]_{xx}-u(b_1u+c_1v) $} &\small {$b_1^2+c_1^2+$}
& \small {$P_t, \, P_x, \, D_1$} & \small {3, 4, 10}\\
  & \small {$v_t = [(d_{21}u+d_{22}v)v]_{xx}-v(b_2u+c_2v)$}   & \small {$b_2^2+c_2^2\not=0$} &
 &\\
%&&& \\
\hline
 %&&& \\
\small {4.} &  \small {$ u_t = [(d_{11}u+d_{12}v)u]_{xx}+a_1u$} &
\small { $a_1 \neq a_2$} & \small {$P_t, \, P_x, \, D_2$} &\small{3, 4}\\
  & \small {$v_t = [(d_{21}u+d_{22}v)v]_{xx}+a_2v$}   &  &
 &  \\
%&&& \\
\hline
% &&& \\
\small {5.} &  \small {$ u_t = [(d_1+v)u]_{xx}+u(a_1-c_1v)  $} &
\small {$d_1^2+d_2^2\not=0$} & \small {$P_t, \, P_x, \, u\pt_u$} &
\small
{1, 2} \\
  & \small {$v_t = [(d_2+d_{22}v)v]_{xx}+v(a_2-c_2v)$}   &  &
 & \\
%&&& \\
\hline
%&&& \\
\small {6.} &  \small {$ u_t = [(d_1+v)u]_{xx}-b_1u^2$} & \small
{$d_1^2+d_2^2\not=0$}  & \small {$P_t, \, P_x, \, D_3$} & \small
{1, 2,} \\
  & \small {$v_t = [(d_2+d_{22}v)v]_{xx}-b_2uv$}   &  &
  & \small {9}  \\
%&&& \\
 \hline
%&&& \\
\small {7.} &  \small {$ u_t = d_{12}[uv]_{xx}+u(a_1-b_1u)  $} & &
\small {$P_t, \, P_x, \, D_4$} & \small
{1, 2,} \\
  & \small {$v_t = [v^2]_{xx}+v(a_2-b_2u)$}   &  &
  & \small
{3, 4}  \\
%&&& \\
 \hline
%&&& \\
\small {8.} &  \small {$ u_t = [(u+d_{12}v)u]_{xx}+au$ } &  \small
{$a\neq0$} & \small {$P_t, \, P_x$,} & \small
{1, 2} \\
  & \small {$v_t = [(u+d_{12}v)v]_{xx}+2av$ }  &  &
 \small{$D_2, \, Q_3$} & \\
%& & & \\
\hline \small {9.} &  \small {$ u_t = [(1+v)u]_{xx}+u(a_1-cv) $} &
% \small { $d\neq0$}
 & \small {$P_t, \, P_x,$} & \small
{1, 2,} \\
  & \small {$v_t = [(1+v)v]_{xx}+v(a_2-cv)$ }  &  &
\small { $u\pt_u, \, Q_4$} &\small
{5, 6, 8}  \\
% &&& \\
\hline \small {10.} &  \small {$ u_t = d_{12}
[uv]_{xx}+u(a_1-c_1v) $} & \small { $a_2\neq0$}  & \small {$P_t,
\, P_x,$}  & \small
{1, 2,} \\
  & \small {$v_t =
  %d_{22}
  [v^2]_{xx}+v(a_2-c_2v)$}   & \small {$c_1\neq c_2$} &
  \small {$u\pt_u, \, Q_5$} & \small
{3, 4} \\
 %&&& \\
\hline
 \small {11.} &  \small {$ u_t = d_{12}
[uv]_{xx}+u(a_1-c_1v)  $} & \small {$c_1\neq c_2$}  & \small
{$P_t, \, P_x,$} & \small
{1, 2,} \\
  & \small {$v_t =
  %d_{22}
  [v^2]_{xx}-c_2v^2$}   &   &
\small {$u\pt_u, \, D_5$} & \small
{3, 4} \\
%&&& \\
\hline
\small {12.} & \small { $ u_t = d_{12} [uv]_{xx}-b_1u^2 $}
& & \small {$P_t, \, P_x$,} & \small
{1, 2, 3,} \\
  & \small {$v_t =
  % d_{22}
  [v^2]_{xx}-b_2uv$}   &  &
  \small {$D_1, \, D_4$} & \small
{4, 9, 10} \\
%&&& \\
\hline
% &&& \\
\small{13.} &  \small{$ u_t =
% d_{12}
[uv]_{xx}+u(a_1-cv)$} & \small{$a_2c\neq0$} & \small{$P_t, \, P_x$,} & \small{1, 2, 7} \\
  & \small{$v_t =
  % d_{12}
  [v^2]_{xx}+v(a_2-cv)$ }  &  &
  \small{$u\pt_u, \, Q_4, \, Q_5$}& \\
% &&& \\
\hline
% &&& \\
\small{14.} & \small{ $ u_t =
%d_{12}
[uv]_{xx}+u(a_1-cv)  $} & \small{$c\neq0$} & \small{$P_t, \, P_x$,} & \small{1, 2, 7} \\
  & \small{$v_t =
  %d_{12}
  [v^2]_{xx}-cv^2$}   &  &
 \small{ $u\pt_u, \, D_5,  \, Q_6$} & \\
%&&& \\
\hline
 %&&& \\
\small{15.} &  \small{$ u_t = d_{12} [uv]_{xx}+a_1u$} &
\small{$a_2\neq0$} & \small{$P_t, \, P_x $,} & \small
{1, 2} \\
  & \small{$v_t =
  %d_{22}
  [v^2]_{xx}+a_2v$}   &\small{$d_{12}\neq 1
  %d_{22}
  $}
   &
 \small{ $u\pt_u, \ D_4, \,  \, Q_5$} & \\
\hline
%&&& \\
\small{16.} &  \small{$ u_t =
%d_{12}
[uv]_{xx}+a_1u$} & \small{$a_2\neq0$}  & \small{$P_t, \, P_x, \,
u\pt_u$} & \small
{1, 2, 8} \\
  & \small{$v_t =
  % d_{12}
  [v^2]_{xx}+a_2v$ }  &  &
\small{ $D_4, \, Q_4, \, Q_5$} & \\
%&&& \\
\hline
\end{tabular}
\end{center}
\end{table}

\begin{remark}  In cases   5, 9, 10, 11, 13---16 of
Table~\ref{tab-1}, the term $a_1 u$ are removable by the
substitution 11 from (\ref{37a}). However, we keep this term because
one has a clear biological
 interpretation (the rate of the birth (or mortality) for species/cells). We also note that some parameters in the systems listed
 in Table~\ref{tab-1} can be reduced to $\pm 1$ by scaling (see substitutions 2 and 10 in (\ref{37a})).
\end{remark}

\begin{remark}
In Tables~\ref{tab-1}--\ref{tab-3}, the following designations for
the Lie symmetry operators are introduced:
\[\ba
D_0=2t\pt_t+x\pt_x;\\
D_1=t\pt_t-(u\pt_u+v\pt_v);\\
D_2=x\pt_x+2(u\pt_u+v\pt_v);\\
D_3=2t\pt_t+x\pt_x-2u\pt_u;\\
D_4=x\pt_x+2v\pt_v;\\
D_5=t\pt_t+a_1tu\pt_u-v\pt_v; \\
D_6=t\pt_t-u\pt_u;\\
Q_1=e^{-at}(\pt_t+a(u-\frac{c}{b}v)\pt_u+2av\pt_v);\\
Q_2=e^{-at}(\pt_t+2au\pt_u+a(\frac{a}{c}-\frac{b}{c}u+v)\pt_v);\\
Q_3=e^{-at}(\pt_t+a(u-d_{12}v)\pt_u+2av\pt_v);\\
Q_4=e^{(a_1-a_2)t}v\pt_u;\\
Q_5=e^{-a_2t}(\pt_t+a_1u\pt_u+a_2v\pt_v);\\
Q_6=e^{a_1t}v\pt_u;\\
R=t\pt_t+\left(\frac{5d_2-4d_1}{3(d_1-d_2)}u+\frac{2d_1-d_2}{3(d_1-d_2)}v-\frac{2d_1-d_2}{3}\right)\pt_u+
\left(\frac{4d_2-5d_1}{3(d_1-d_2)}v+
\frac{d_1-2d_2}{3(d_1-d_2)}u+\frac{d_1-2d_2}{3}\right)\pt_u;\\
 Z_1=\frac{e^{x}}{u-v}\left(\pt_u
 -\pt_v\right), \
 Z_2=\frac{e^{-x}}{u-v}\left(\pt_u
 -\pt_v\right);\\
  Z_3=\frac{\cos x}{u-v}\left(\pt_u
 -\pt_v\right), \
 Z_{4}=\frac{\sin x}{u-v}\left(\pt_u
 -\pt_v\right);\\
 Z_{5}=\frac{x}{u-v}\left(\pt_u
 -\pt_v\right), \
 Z_{6}=\frac{1}{u-v}\left(\pt_u
 -\pt_v\right).
 \ea \]
\end{remark}

Now we examine system (\ref{2}) under the condition
 $d_1=d_2=d_{11}=d_{22}=0$, which reduces the system  to the form
\be\label{109} \ba
 \medskip
 u_t = d_{12}[u v]_{xx}+u(a_1-b_1u-c_1v),\\
 v_t = d_{21}[u v]_{xx}+v(a_2-b_2u-c_2v).
 \ea
  \ee
Assuming  $d_{12}d_{21}\neq 0$, we transform (\ref{109}) to the form
 \be\label{111} \ba
 \medskip
 u_t = [u v]_{xx}+u(a_1-b_1u-c_1v),\\
 v_t = [u v]_{xx}+v(a_2-b_2u-c_2v)
 \ea
  \ee
  by the simple transformation
\be\label{110}
 u^*=d_{21}u,\quad
 v^*= d_{12}v
  \ee
(in system (\ref{111}), stars next to $u$ and $v$ are skipped). In the case of
the coefficients arising in system (\ref{111}) the system of DEs
(\ref{10})---(\ref{26}) takes an essentially different form. As a
result, equation (\ref{35}) is not obtainable. An analysis of the
system of DEs is omitted here. The final result can be formulated as
follows.

\begin{theo}\label{th-2}
All possible  MAIs (up to representations generated by
transformations of the form (\ref{112})) of system (\ref{111}) with
non-vanish reaction term(s) are presented in Table~\ref{tab-2}. Any
other system of the form (\ref{111}), which is invariant w.r.t. the
three- and higher-dimensional Lie algebra is reduced by a
substitution of the form \be\label{112} \ba
\medskip
1. \ t^*=\frac{1}{a}e^{a t}, \quad u^*=e^{-a t}u, \quad
v^*=e^{-a t}v.\\
2. \ t^*=bt, \ x^*=\sqrt{b}x, \, b>0 \ea \ee ($a$ and $b$ are
arbitrary non-zero constants) to one of those given in
Table~\ref{tab-2}.
\end{theo}

\begin{table}
\caption{MAIs of  system (\ref{111})} \label{tab-2}
\begin{center}
\begin{tabular}{|c|c|c|c|c|}
\hline
\small  &\small {Systems}  & \small {Restrictions} & \small {Basic operators  } & \small {Substitution}   \\
 & &  & \small {of MAI} &  \small {from (\ref{112})}   \\
\hline {1.} &  \small {$ u_t = [uv]_{xx}-u(b_1u+c_1v) $} &\small
{$b_1^2+c_2^2\not=0$}  &
\small {$P_t, \, P_x, \, D_1$} & \small {$1$}\\
  & \small {$v_t = [uv]_{xx}-v(b_2u+c_2v)$}   & &
 &  \\
%&&& \\
\hline
 %&&& \\
\small {2.} &  \small {$ u_t = [uv]_{xx}+a_1u$} &
\small { $a_1 \neq a_2$} & \small {$P_t, \, P_x, \, D_2$} &\\
  & \small {$v_t = [uv]_{xx}+a_2v$}   &  &
 &  \\
%&&& \\
\hline \small {3.} &  \small {$ u_t = [uv]_{xx}-uv$} &
& \small {$P_t, \, P_x, \, D_1,$} & \small {$1, \ 2$}\\
  & \small {$v_t = [uv]_{xx}-uv$}   &  &
 \small {$Z_1, \, Z_2$} &  \\
%&&& \\
\hline \small {4.} &  \small {$ u_t = [uv]_{xx}+uv$} &
 & \small {$P_t, \, P_x, \, D_1,$} & \small {$1, \ 2$}\\
  & \small {$v_t = [uv]_{xx}+uv$}   &  &
 \small {$Z_3, \, Z_{4}$} &  \\
%&&& \\
\hline
\end{tabular}
\end{center}
\end{table}

Finally, we examine   system (\ref{2})
%with $d^2_{12}+d^2_{21}\not=0$ and
without reaction terms, i.e., the cross-diffusion system  \be\label
{113}
 \ba
 \medskip
 u_t = [(d_1+d_{11}u+d_{12}v)u]_{xx},\\
 v_t = [(d_2+d_{21}u+d_{22}v)v]_{xx}.
 \ea
  \ee

It is worth to note that system (\ref{113}) with arbitrary
coefficients is invariant under the three-dimensional trivial
algebra spanned by the basic operators  $P_t, \,  P_x$ and $D_0.$
All possible extensions of this trivial algebra admitting by
(\ref{113}) are presented in the following statement.

 \begin{theo}\label{th-3}
All possible  MAIs (up to equivalent representations generated by
transformations of the form (\ref{114})) of system (\ref{113}) are
presented in Table~\ref{tab-3}. Any  other system of the form
(\ref{113}) which is invariant w.r.t. the four- and
higher-dimensional Lie algebra is reduced by a substitution of the
form \be\label{114} \ba t^*=\alpha_1t, \
u^*=\al_{2}+\al_{3}u+\al_{4}v, \ v^*=\al_{5}+\al_{6}u+\al_{7}v\ea
\ee to one of those given in Table~\ref{tab-3} (constants $\al_{i}$
are determined by the system in question).
\end{theo}

\begin{table}
\caption{MAIs of  system (\ref{113}) } \label{tab-3}
\begin{center}
\begin{tabular}{|c|c|c|c|c|}
\hline
\small  &\small {Systems}  & \small {Restrictions} & \small {Basic operators  } & \small {Substitution}   \\
 & &  & \small {of MAI} &  \small {from (\ref{115})}   \\
\hline
 %&&& \\
\small {1.} &  \small {$ u_t = [(d_{11}u+d_{12}v)u]_{xx}$} &
$(d_{11}-d_{21})^2+
(d_{12}-d_{22})^2\neq0$ & \small {$P_t, \, P_x, \, D_0, \, D_1$} & \small{$1, \ 2$}\\
  & \small {$v_t = [(d_{21}u+d_{22}v)v]_{xx}$}   & $d^2_{11}+d^2_{21}\neq0$ &
 &  \\
&&$d^2_{12}+d^2_{22}\neq0$& &\\
\hline
% &&& \\
\small {2.} &  \small {$ u_t = [(d_1+d_{11}u)u]_{xx}$} & \small
{$d_{11}\not=1$}  & \small {$P_t, \, P_x, \, D_0, \, v\pt_v$} &  \small{$1, \ 3$}\\
  & \small {$v_t = [(d_2+u)v]_{xx}$}   & \small
{$d_{1}\not=2d_{2}d_{11}$}  &
   &   \\
%&&& \\
 \hline
 %&&& \\
\small {3.} &  \small {$ u_t = [(d_1+u+v)u]_{xx}$} & \small
{$d_{1}\not=d_2$}  & \small {$P_t, \, P_x, \, D_0, \, R$} &  \small{$4$}\\
  & \small {$v_t = [(d_2+u+v)v]_{xx}$}   &  &
 &   \\
%&&& \\
 \hline
%&&& \\
\small {4.} &  \small {$ u_t = d_{11}[u^2]_{xx}$} & \small
{$d_{11}\not=1$}  & \small {$P_t, \, P_x, \, D_0$} & \small{$1, \ 7$}\\
  & \small {$v_t = [uv]_{xx}$}   &  &
 \small {$v\pt_v, \, D_6$}  &   \\
%&&& \\
 \hline
 %&&& \\
\small {5.} &  \small {$ u_t = [(1+u)u]_{xx}$} & & \small {$P_t, \,
P_x,  \, D_0,$}
&  \small{$1, \ 5, \ 6$}\\
  & \small {$v_t = [(1+u)v]_{xx}$}   &  &
 \small {$v\pt_v, \, u\pt_v$}  &   \\
%&&& \\
 \hline
%&&& \\
\small {6.} &  \small {$ u_t = [u^2]_{xx}$} & & \small {$P_t, \,
P_x, \, D_0, \, D_6,$} & \small{$1, \ 5, \ 7$}\\
  & \small {$v_t = [uv]_{xx}$}   &  &
 \small {$v\pt_v, \,   u\pt_v$}  &  \small{$$} \\
%&&& \\
 \hline
%&&& \\
\small {7.} &  \small {$ u_t = [uv]_{xx}$} &
 & \small {$P_t, \, P_x, \, D_0,$} & \small{$4$}\\
  & \small {$v_t = [uv]_{xx}$}   &  &
 \small {$D_1, \, Z_{5}, \, Z_{6}$} &  \\
%&&& \\
\hline
\end{tabular}
\end{center}
\end{table}

In Table~\ref{tab-3}, the following substitutions are used:
\be\label{115}\ba
1. \quad  u^*=v, \quad v^*=u;\\
2. \quad  u^*=u, \quad v^*=v+\frac{d_1}{d_{12}};\\
3. \quad u^*=d_{21}u, \quad v^*= v;\\
4. \quad u^*=d_{11}u, \quad v^*= d_{12}v;\\
5. \quad t^*=d_1t, \quad d_1u^*=d_{11}u+d_{12}v, \quad v^*= v;\\
6. \quad t^*=(2d_2-d_1)t, \quad (2d_2-d_1)u^*=d_{11}u+d_{1}-d_2, \quad v^*= v;\\
7. \quad u^*=d_1+d_{21}u, \quad v^*= v. \ea \ee

It should be stressed that MAIs of the nonlinear systems
\be\label{1-4} \ba
 u_t = [uv]_{xx}+buv,\\
 v_t =[uv]_{xx}+buv
 \ea\ee ($b=\pm1$ in the cases 3 and 4 of Table~\ref{tab-2} and $b=0$ in the case 7 of Table~\ref{tab-3})
contain the Lie symmetry operators, which are \textbf{nonlinear}
w.r.t. to the dependent variables $u$ and $v$. It is new property of
RD systems with cross-diffusion, which not occurs for the standard
RD systems (see Lie symmetries in \cite{ch-king, ch-king03,
ch-king4} and  papers  cited therein).

\begin{remark} System (\ref{1-4}) can be simplified as
follows \be\label{1-5} \ba
 \medskip
 u_t = [(u-w)u]_{xx}+bu(u-w),\\
 w_t = 0
 \ea
  \ee
  by the transformation $w=u-v$. The symmetry operators mentioned above will be transformed to those, which are again nonlinear w.r.t. the depended variable(s).
  \end{remark}

\section{\bf Example of exact solutions} \label{sec-3}

 Here we consider the systems  \be\label{3-1} \ba
 u_t = [uv]_{xx}-uv,\\
 v_t =[uv]_{xx}-uv
 \ea
  \ee and \be\label{3-2} \ba
 u_t = [uv]_{xx}+uv,\\
 v_t =[uv]_{xx}+uv
 \ea
  \ee from cases 3 and 4 of Table~\ref{tab-2}, which admit the most nontrivial
  algebras of invariance and are  important subcases (up to local transformations) of the SKT system (\ref{2}).

Let us construct exact solutions of the nonlinear systems (\ref{3-1}) and
(\ref{3-2}) using the Lie symmetries $X_1=\lambda_1Z_1+\lambda_2Z_2$
 and $X_2=\lambda_1Z_3+\lambda_2Z_4$, respectively (hereafter $\lambda_1$ and $\lambda_2$ are arbitrary constants,
 $\lambda^2_1+\lambda^2_2\neq0$).

It is well-known that Lie symmetry operators with a complicated
structure can be used for finding nontrivial solutions from the very
simple ones (probably the first  examples for RD systems were
presented in \cite{ch-1988}, see also \cite{ch-king}). Here we show
how it can be realize for the nonlinear system (\ref{3-1})
%using the operator $X_1$
(exact solutions of (\ref{3-2}) can be obtained in a quite similar way).

Operator $X_1$ generates the one-parameter Lie group: \be\label{3-3}\ba
\medskip
u^*
=\frac{u+v}{2}+\frac{1}{2}\sqrt{(u-v)^2+4p(\lambda_1e^{x}+\lambda_2e^{-x})}\,,
\\
v^*=\frac{u+v}{2}-\frac{1}{2}\sqrt{(u-v)^2+4p(\lambda_1e^{x}+\lambda_2e^{-x})},
\ea\ee when $u\geq v$; \be\label{3-4}\ba \medskip
u^*=\frac{u+v}{2}-\frac{1}{2}\sqrt{(u-v)^2+4p(\lambda_1e^{x}+\lambda_2e^{-x})}\,,
\\
v^*=\frac{u+v}{2}+\frac{1}{2}\sqrt{(u-v)^2+4p(\lambda_1e^{x}+\lambda_2e^{-x})}\,,
\ea\ee  when $u< v$ (here $p$ is an arbitrary parameter).

In order to construct a nontrivial solution $(u^*,v^*)$ of system (\ref{3-1}), one needs to know a simple exact solution, which can be easily derived by setting
$u_x=v_x=0.$ It means that  the ODE system  \be\nonumber\ba
 u_t = -uv,\\
 v_t =-uv
 \ea\ee is obtained instead of (\ref{3-1}) and its general solution is
\be\label{3-5}\ba
 u=\frac{\alpha_1e^{\alpha_1t}}{\alpha_2+e^{\alpha_1t}},\\
 v=-\frac{\alpha_1\alpha_2}{\alpha_2+e^{\alpha_1t}},
 \ea\ee
where $\alpha_1$  and $\alpha_2$ are arbitrary constants.
Thus, solution (\ref{3-5}) of system (\ref{3-1}) can be  generalized  via transformations (\ref{3-3})--(\ref{3-4}) to the five-parameter family of solutions
 \be\label{3-6}\ba \medskip
u(t,x)=\frac{\alpha_1e^{\alpha_1t}-\alpha_1\alpha_2}{2(\alpha_2+e^{\alpha_1t})}\pm\frac{1}{2}\sqrt{\alpha_1^2+4p(\lambda_1e^{x}+\lambda_2e^{-x})}\,,
\\
v(t,x)=\frac{\alpha_1e^{\alpha_1t}-\alpha_1\alpha_2}{2(\alpha_2+e^{\alpha_1t})}\mp\frac{1}{2}\sqrt{\alpha_1^2+4p(\lambda_1e^{x}+\lambda_2e^{-x})}\,.
\ea\ee

Using the same algorithm, the following family of  exact solutions of system
(\ref{3-2}) was  derived: \be\label{3-7}\ba \medskip
u(t,x)=\frac{\alpha_1+\alpha_1\alpha_2e^{\alpha_1t}}{2(1-\alpha_2e^{\alpha_1t})}\pm\frac{1}{2}\sqrt{\alpha_1^2+4p(\lambda_1\cos
x+\lambda_2\sin x)}\,,
\\
v(t,x)=\frac{\alpha_1+\alpha_1\alpha_2e^{\alpha_1t}}{2(1-\alpha_2e^{\alpha_1t})}\mp\frac{1}{2}\sqrt{\alpha_1^2+4p(\lambda_1\cos
x+\lambda_2\sin x)}\,. \ea\ee

Notably,  one may set $\alpha_2=\pm 1$ in the solutions
(\ref{3-6})--(\ref{3-7}) without losing a generality while the
case $\alpha_2=0$ leads to steady-state solutions.

\begin{remark} All possible  steady-state solutions of
the nonlinear systems (\ref{3-1}) and (\ref{3-2}) can be easily derived. As a result one obtains  \be\label{3-13}\ba
 u(x)=\frac{f(x)}{g(x)},\quad
 v(x)=g(x)\not= 0,\\
 u(x)=h(x), \quad v(x)=0, \\
 u(x)=0, \quad v(x)=h(x),
 \ea \ee where $g(x)$ and $h(x)$  are  arbitrary smooth  functions, while the function
 $f(x)$ is the general solution of the linear ODE $f''\mp f=0 $. Obviously, transformations (\ref{3-3})--(\ref{3-4})
 do not generate  new solutions from  (\ref{3-13}).
 \end{remark}

It should be emphasized that  each solution of the form  (\ref{3-7})  satisfy
the zero flux conditions on a correctly-specified space interval. For example, the exact solution with $\lambda_2=0$ satisfy the zero
 Neumann conditions
 \be\label{3-9} u_x|_{x=0}=0, \ v_x|_{x=0}=0,  \
u_x|_{x=\pi}=0, \ v_x|_{x=\pi}=0 \ee at the interval
$(0,\pi)$. This property  is important for possible applications because zero
flux at boundary is a typical requirement for biologically motivated
models.
It should be also noted that
systems (\ref{3-1}) and (\ref{3-2}) are
two canonical forms of a more general system. In fact, the
substitution \be\nonumber t^*=\frac{\ln t}{ab}, \
x^*=\frac{x}{\sqrt{b}} \ (a\not=0, b>0), \ u^*=\frac{atu}{d_2}, \
v^*=\frac{atv}{d_1} \ee  transforms  systems (\ref{3-1}) and (\ref{3-2})
to the form \be\label{3-8} \ba
 u^*_{t^*} = d_1[u^*v^*]_{x^*x^*}+u^*(ab+b_1v^*),\\
 v^*_{t^*} =d_2[u^*v^*]_{x^*x^*}+v^*(ab+b_2u^*),
 \ea\ee where $b_1=\pm bd_1,$  $b_2=\pm
bd_2.$ Now one realizes that system (\ref{3-8}) involves the logistic type terms, which are very common in the mathematical biology models.

% Зауважимо, що система \be\label{3-8} \ba
 %u_t = [uv]_{xx}\pm uv,\\
 %v_t =[uv]_{xx}\pm uv
 %\ea\ee

  Finally, we present an example for deriving  exact solutions via the most common procedure, which is often called the Lie symmetry reduction. Let us take a linear combination of the
   Lie symmetries   $P_x, \ Z_3$ and $Z_4$, i.e.,
\be\label{3-10}
X=\pt_x+\frac{\lambda_1\cos x+\lambda_2 \sin
x}{u-v}(\pt_u-\pt_v).\ee
Obviously,  (\ref{3-10}) is again a Lie symmetry, which produces
the  ansatz \be\label{3-11} \ba
 u= \frac{\varphi_1(t)\pm\sqrt{\varphi_1^2(t)+4\varphi_2(t)+4(\lambda_1\sin x-\lambda_2
 \cos x)}}{2},\\
 v=\frac{\varphi_1(t)\mp\sqrt{\varphi_1^2(t)+4\varphi_2(t)+4(\lambda_1\sin x-\lambda_2
 \cos x)}}{2},
 \ea\ee where $\varphi_1(t)$ and $\varphi_2(t)$ are
new unknown functions. Formally speaking,  one should take the upper signs if $u\geq v$, otherwise the lower signs should be used, however it  is not essential because system (\ref{3-2}) is invariant under the discrete transformation $u \to v, \ v \to u$.

 Ansatz (\ref{3-11}) reduces the nonlinear system (\ref{3-2}) to the ODE system
\be\label{3-12*} \ba
\varphi_1'+2\varphi_2=0,\\
\varphi_1\varphi_1'+2\varphi_2'=0. \ea\ee
In contrast to (\ref{3-2}), system  (\ref{3-12*}) is integrable because  is equivalent  to the system
\be\label{3-12} \ba
\varphi_2=-\frac{1}{2}\varphi_1',\\
\varphi_1'=\frac{1}{2}\varphi_1^2+\beta, \ea\ee
in which the general solution of  the second equation is well-known.
Thus, having   the general solution of the reduced system (\ref{3-12}) and using ansatz (\ref{3-11}), three different  solutions  (depending on the sign of the constant $\beta$) of system (\ref{3-2}) were found:
\be\label{3-14} \ba
 u= -\frac{1}{t}\pm\sqrt{\lambda_1\sin x-\lambda_2
 \cos x},\\ \medskip
 v=-\frac{1}{t}\mp\sqrt{\lambda_1\sin x-\lambda_2
 \cos x},\\
 u= \alpha_1\tan(\alpha_1t)\pm\sqrt{\lambda_1\sin x-\lambda_2
 \cos x-\alpha_1^2},\\ \medskip
 v=\alpha_1\tan(\alpha_1t)\mp\sqrt{\lambda_1\sin x-\lambda_2
 \cos x-\alpha_1^2},\\
 u= \alpha_1\frac{1+\alpha_2e^{2\alpha_1t}}{1-\alpha_2e^{2\alpha_1t}}\pm\sqrt{\lambda_1\sin x-\lambda_2
 \cos x+\alpha_1^2},\\
 v=\alpha_1\frac{1+\alpha_2e^{2\alpha_1t}}{1-\alpha_2e^{2\alpha_1t}}\mp\sqrt{\lambda_1\sin x-\lambda_2
 \cos x+\alpha_1^2},
 \ea\ee where $\alpha_1=\pm \sqrt{\frac{|\beta|}{2}}$  and  $\alpha_2$ is
 an arbitrary constant. Notably, the last solution from (\ref{3-14}) is a particular case of  (\ref{3-7}).

 %TO ADD about periodic solution !!!!!

%\newpage

\section{\bf Conclusions} \label{sec-4}

In this paper, the
 %complete description of
 Lie symmetry classification problem  of the Shigesada--Kawasaki--Teramoto system is completely  solved. Solution of this problem was initiated in \cite{ch-my2008},
however, the result derived therein is not complete because
essential restriction on coefficients were applied, as a result all the symmetries derived in \cite{ch-my2008} can be extracted from Table 1. Here it is
proved that the SKT system (\ref{2})  admits a wide range of   Lie
symmetries depending on 12  coefficients arising in the system (see
Tables~\ref{tab-1}--\ref{tab-3}). From the applicability point of
view, the most interesting systems occur in Tables~\ref{tab-1}
and~\ref{tab-2}. For example, the systems listed in cases 1--4 have
a quite general structure and one may expect that some SKT models
with the correctly-specified coefficients (which are chosen from
experimental data) are equivalent to these systems. One may note
that several systems in Table~\ref{tab-1} (cases 5--7, 9--16)
contain the cross-diffusion term in the first equation only. Such
systems occurs when one of the cross-diffusion coefficients is much
larger  than the other (see the pioneering paper
\cite{mimura-et-all-84} for detail). They are called
triangular   and are extensively studied  during the last decade
 (see  \cite{des-tre-2015} and references  therein).

From the Lie symmetry point of view the most interesting systems
occur in Tables~\ref{tab-2} and~\ref{tab-3}. In particular, the Lie
symmetry operators with highly unusual structure are unveiled for
the nonlinear systems listed in cases 3--4 of Table~\ref{tab-2} and
in case  7 of Table~\ref{tab-3}. In fact, operators $Z_1, \dots,
Z_6$ are \textbf{nonlinear w.r.t.  the dependent variables $u$ and
$v$} because they involve  coefficients of the form $
\frac{f(x)}{u-v} $ ($f(x)$ is a correctly-specified function). To
the best of our knowledge, this is the first time when nonlinear Lie
symmetry operators are found for RD systems. In the case of RD
systems without cross-diffusion, all possible Lie symmetry operators
are known (see \cite{ch-king4} and references therein) and they are
always linear w.r.t.  the dependent variables. In the case of RD
systems involving cross-diffusion, there is no a complete
description of all possible Lie symmetry operators at the present
time. However, the results obtained  in \cite{niki-05} (the case of
constant cross-diffusion), \cite{ch-wil-96} (power-law coefficients
of cross-diffusion),\cite{tor-tra-val-96} (diffusion and
cross-diffusion in the first equation and no any diffusion in the
second) and \cite{serov-et-al-15} (Galilei-invariant  systems with
cross-diffusion) show that all Lie symmetries of RD systems found
therein are linear w.r.t. to unknown functions. We foresee that new
nonlinear Lie symmetry operators will be found for suitable
generalizations  of   the SKT system (\ref{2}). Notably, nonlinear
Lie symmetry operators do not occur in the case of any single RD
equation \cite{dor}, however it was recently established that the RD
equation with a correctly-specified  gradient-dependent diffusivity
and an arbitrary reaction term admits such operators and is
linearizable \cite{ch-ki-ko-16}  (see Theorem 1 therein).

Finally, the Lie symmetry classification is  applied for finding exact
solutions of the  nonlinear systems, which are invariant under the operators mentioned above. Our purpose was to show how highly  non-trivial Lie symmetries  generate exact solutions, which may be useful in applications. In particular, we have  shown that some exact solutions satisfy zero
flux  boundary conditions, which are    typical requirements for solutions of biologically motivated models.

\end{document}